\documentclass{elsart}
\usepackage{graphicx}
\usepackage{amssymb}

\begin{document}

\begin{frontmatter}

\title{Effects of parametric noise on a nonlinear oscillator}

\author{Kirone Mallick}
\address{Service de Physique Th\'eorique, Centre d'\'Etudes de Saclay,\\
 91191 Gif-sur-Yvette Cedex, France}
\ead{mallick@spht.saclay.cea.fr}
\author{Philippe Marcq}
\address{Institut de Recherche sur les Ph\'enom\`enes Hors \'Equilibre,
 Universit\'e de Provence,
 49 rue Joliot-Curie, BP 146, 13384 Marseille Cedex 13, France}
\ead{marcq@irphe.univ-mrs.fr}

\begin{abstract}
  We study a model of a  nonlinear oscillator
 with a random frequency  and 
  derive  the  asymptotic
 behavior of the probability distribution function when
 the noise is white. 
 In the small damping
  limit,  we show that the  physical observables grow algebraically
 with time before the dissipative time scale is reached,  and calculate the
  associated anomalous diffusion exponents.  In the case of
  colored  noise, with a nonzero  but arbitrarily small 
correlation time, the   characteristic  exponents are  modified.
 We determine their values thanks to a self-consistent Ansatz.
 \end{abstract}

\end{frontmatter}

 \section{Introduction}

 A simple  model of a nonlinear stochastic system is obtained by including
 a non-quadratic confining potential 
 in the classical Langevin equation  \cite{wax,vankampen}.
 In this work,  we address the problem of a nonlinear oscillator
 with random  linear frequency. 
Due to  the continuous injection
 of  energy by the random force, 
   observables   such as 
  the oscillator's mechanical energy, position and 
  velocity, grow algebraically with time in  the absence of dissipation.
  In  Section 2,    we  calculate
  the associated   growth exponents in the case of Gaussian
 white noise,   derive 
 the   long time asymptotic behavior 
  of the probability distribution  function
 (P.D.F.) in phase  space  and match it with the  (non-Gibbsean)
 stationary P.D.F.  in presence of small dissipation. 
 In Section 3, we prove 
 that if the noise has a non-vanishing 
 correlation time, the dynamics of the nonlinear oscillator
 is drastically modified:  colored  noise
  yields anomalous diffusion exponents equal
  to half the values  found for white noise.

\section{White multiplicative noise}

 We   consider  a nonlinear oscillator 
 of amplitude $x(t)$, submitted  to  parametric noise
 and confined by  a potential ${\mathcal U}(x)$.
  For $ |x| \rightarrow \infty$, a suitable rescaling permits us to
  write 
  $   {\mathcal U}   \sim \frac{ x^{2n}}{2n} $
    with  $ \, n \ge  1 .$ Thus,  for large amplitudes,
 the stochastic differential  equation that governs the system becomes 
 \begin{equation}
 \frac{\textrm{d}^2 }{\textrm{d} t^2}x(t) 
    + \gamma  \frac{\textrm{d} }{\textrm{d}t }x(t)
  + x(t)^{2n-1}  = \xi(t)  x   \,,
  \; \hbox{ with } \;
 \langle \xi(t)  \xi(t') \rangle  =  
    {\mathcal D} \, \delta( t - t')  ,
 \label{dynwhite}
\end{equation}
where  $\gamma$ is  the damping rate and    
   $\xi(t)$ is a Gaussian white noise  of zero mean-value and 
 of autocorrelation ${\mathcal D}$. 
 This   equation is 
  interpreted according to the rules of  Stratonovich Calculus
  \cite{vankampen}.
 For a  quadratic  potential ({\it i.e.} $ n =1$)
 and for small $\gamma$, 
 the energy of the oscillator  grows exponentially with time
  \cite{hansel}.
 The effect  of nonlinear restoring forces must be taken 
 into account  to avoid this  exponential  amplification.

 We shall  analyse the motion of the  nonlinear stochastic oscillator
    following the method
 explained in \cite{philkir1,philkir2}. Defining the energy
 and  the angle variables as
\begin{equation}
  E =  \frac{1}{2}\dot x^2 + \frac{1}{2n} x^{2n} \, \,\, ,
  \;\;\; \hbox{ and } \;\;\;   \phi  = 
 \frac{ \sqrt{n}} { (2n)^{1/2n} } \int_0^ { {x}/{E^{{1}/{2n}}} }
   \frac{{\textrm d}u}{\sqrt{ 1 -  \frac{u^{2n}}{2n}}}    \, , 
\label{defphi}
\end{equation}
 we  transform  the coordinates in phase space 
 from position and velocity  to   energy and angle.   
 Introducing  an  auxiliary variable $\Omega$, with   
 $ \Omega =   (2n)^{ \frac{n+1}{2n} } \,  E^{\frac{n-1}{2n}}  \, , $ 
 we rewrite Eq.~(\ref{dynwhite}) as 
   \begin{eqnarray}
     \dot \Omega  &=&   -\gamma \; \frac{n -1 } {  (2n)^{\frac{1}{n}}} \;
  {\mathcal S}_n'(\phi)^2 \; \Omega  \, + \, 
  (n -1) \, {\mathcal S}_n(\phi)
 {\mathcal S}_n'(\phi) \; \xi(t)   \label{evoln1}
   \,  ,   \\    \dot\phi  &=&  \gamma  \;  \frac{ {\mathcal S}_n(\phi)
  {\mathcal S}_n'(\phi) }{  (2n)^{\frac{1}{n}}}   \, + \, 
 \frac{\Omega}{ (2n)^{\frac{1}{n}}}
  - \frac{{\mathcal S}_n(\phi)^2}{\Omega}  \, \xi(t)  \, ,
    \label{evoln}
   \end{eqnarray}
 where  the hyperelliptic function ${\mathcal S}_n$ satisfies 
 ${\mathcal S}_n(\phi) = x/ E^{1/(2n)}$.
 These equations  are rigorous
 and have been derived without any
 hypothesis on the noise term.  In the small damping  limit 
  $\gamma \rightarrow  0$, we deduce from Eqs.~(\ref{evoln1},\ref{evoln})
   that   $\Omega \sim t$ and     $\phi \sim t^2$, {\it i.e.}
 $\phi$  is  a fast  variable as compared to $\Omega$.
  Averaging
 the dynamics  over    $\phi$  yields
  the statistical equipartition 
 identities \cite{philkir1}:
 \begin{equation}
   \label{eq:equip}
   \langle  E   \rangle  =   \frac{n+1}{2n} \, \langle \dot x^2   \rangle 
\hspace*{0.5cm} \mathrm{and} \hspace*{0.5cm}
\langle  x^{2n}   \rangle  =   \langle \dot x^2   \rangle  \,.
 \end{equation}

\begin{figure}[t]
\centerline{\includegraphics*[width=0.8\textwidth]{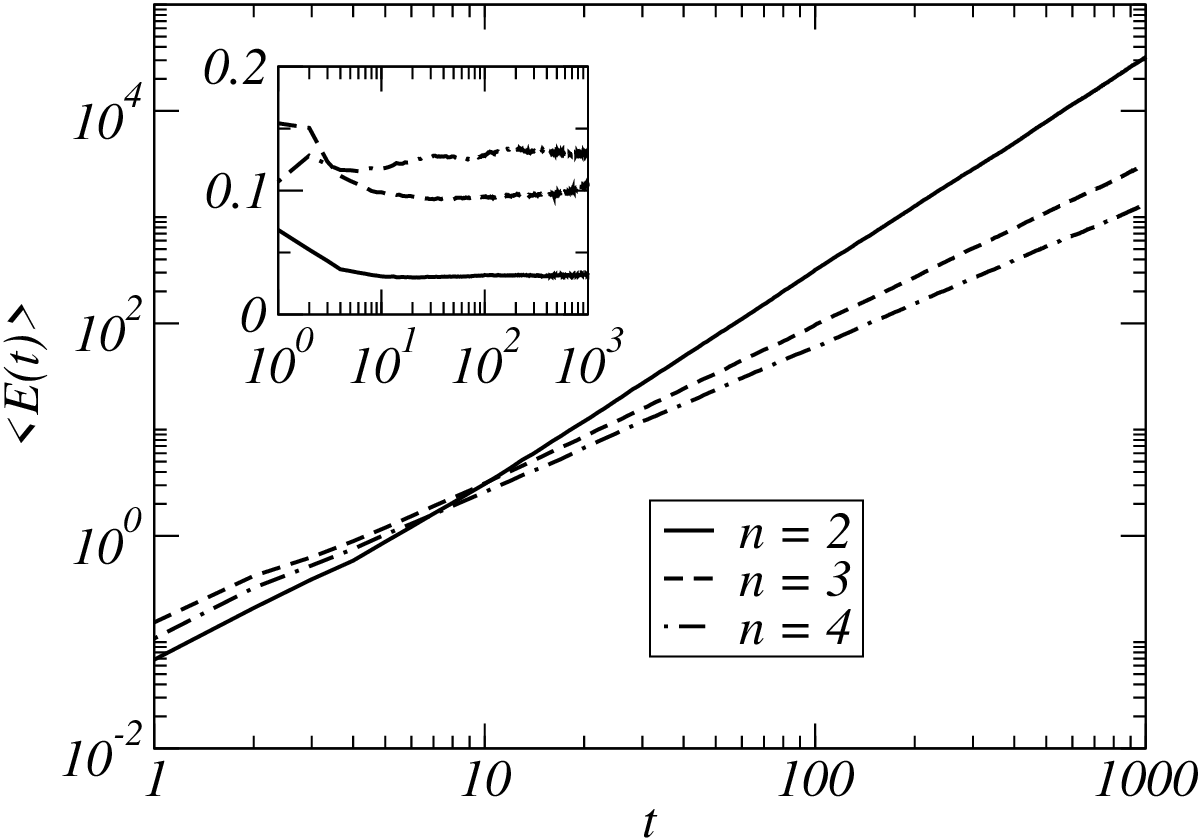}}
 \caption{\label{fig:scaling:white} 
Scaling behavior of $\langle E(t) \rangle$ for white,
multiplicative noise. Inset: the ratio $\langle E \rangle/t^{n/(n-1)}$
is plotted vs. time.
Eq.~(\ref{dynwhite}) is integrated numerically for $\gamma = 0$, 
${\mathcal D} = 1$, 
with a timestep $\delta t$, and averaged over $10^4$ realizations for
$n=2$, $\delta t = 5 \, 10^{-4}$;
$n=3$, $\delta t = 5 \, 10^{-4}$;
$n=4$, $\delta t = 10^{-4}$.
}
\end{figure}

 This   averaging procedure \cite{strato,lindenberg}
 allows us to derive  
  a closed  equation  for  the stochastic
 evolution  of   the slow variable  $\Omega.$ 
 We  start  with   the Fokker-Planck equation 
 for  the P.D.F. $P_t(\Omega, \phi)$
  associated  with the system (\ref{evoln1},
\ref{evoln}),  and  average it under the hypothesis that 
 the probability density 
becomes uniform in $\phi$  when $ t \to \infty$. 
 Hence,  the reduced probability density $\tilde{P}_t(\Omega)$ satisfies
\begin{eqnarray}
   \partial_t {\tilde P}  = \gamma \; \frac{n-1}{n+1} \;
  \partial_{\Omega}\left(\Omega {\tilde P}   \right) \,\, + \,\, 
     \frac { {\tilde{\mathcal D} }}{2} 
\left(  \partial_{\Omega}^2 {\tilde P} - \frac{2}{n-1} \, \partial_{\Omega}
      \frac{{\tilde P}}{\Omega}  \right)  \,.
\label{avFP}
\end{eqnarray} 
 This averaged 
Fokker-Planck equation  is associated  with
 the following  effective
 Langevin dynamics for the variable $\Omega$,
\begin{equation}
   \dot\Omega = \frac{{\tilde {\mathcal D} }} { n-1 } \;
 \frac{1} {\Omega}   -  \gamma  \;
 \frac{n-1}{n+1} \;  \Omega       + \tilde{ \xi} (t)  
\label{Langeff} 
\end{equation}  
where the effective  Gaussian  white noise $\tilde{ \xi}$
 has an amplitude ${\tilde {\mathcal D} }$ given by
\begin{equation}
 {\tilde{\mathcal D}}     = {\mathcal D} \;
  (2n)^{\frac{2}{n} } \;  \frac { (n-1)^2}{n+1}   \,  \,
     \frac{\Gamma \left(\frac{3}{2n} \right)
 \; \Gamma \left(\frac{3n+1}{2n} \right)} 
  {\Gamma \left(\frac{1}{2n} \right) \; 
 \Gamma \left(\frac{3n+3}{2n} \right)} .
\label{Deff} 
\end{equation}  
If  the dissipation rate
  $\gamma$ is taken to be zero, 
 the   solution of Eq.~(\ref{avFP}) 
 is  \cite{philkir1}:
 \begin{equation}
   {\tilde P}_t(E) =
  \frac{1}{  \Gamma \left(\frac{n + 1}{2 (n-1)}\right)} \,
\frac{n-1}{n E} \,
\left( \frac{(2n)^{\frac{n+1}{n}} \; E^{\frac{n-1}{n}}}
{2 {\tilde{\mathcal D}} t}\right)^{\frac{n+1}{2 (n-1)}}
\exp \left\{ - 
\frac{(2n)^{\frac{n+1}{n}} \; E^{\frac{n-1}{n}}}
{2 {\tilde{\mathcal D}}  t}
 \right\} \, .
\label{pdfmult}
\end{equation} 
 From this P.D.F.,  we    derive  analytical expressions
 for statistical averages of physical observables \cite{philkir1}.
 The leading  scaling behaviour is given by
 \begin{equation}
    E    \sim   ({\mathcal D}t)^{\frac{n}{n-1}}  \,,  \;\;\;\;\;\;
    x \sim  ({\mathcal D}t)^{\frac{1}{2(n-1)}} \,,   \; \;\; 
   \; \;\;  \dot{x} \sim   ({\mathcal D}t)^{\frac{n}{2(n-1)}}  \,.
 \label{scalmult} 
 \end{equation} 
 These formulae are 
  in excellent agreement with numerical  simulations \cite{philkir1}.
We show in Fig.~\ref{fig:scaling:white}  that the averaged energy
 scales like  $t^{\frac{n}{n-1}}$  as predicted.  The amplitude  $x$ 
 undergoes an anomalous  diffusion   with an exponent
 that diverges as  $n \to 1$, which is 
 consistent   with the exponential growth
 of the linear oscillator. 

 In the presence of dissipation,
the stationary  solution of the  Fokker-Planck equation (\ref{avFP})  is
 \begin{equation}
   {\tilde P}_{\rm{st}}(E) =
  \frac{1}{  \Gamma \left(\frac{n + 1}{2 (n-1)}\right)} \,
\frac{n-1}{n E} \,
\left( \frac{(2n)^{\frac{n+1}{n}} \; E^{\frac{n-1}{n}}}
{ \frac{n + 1}{\gamma (n-1)} 
   {\tilde{\mathcal D}} }\right)^{\frac{n+1}{2 (n-1)}}
\exp \left\{ - 
\frac{(2n)^{\frac{n+1}{n}} \; E^{\frac{n-1}{n}}}
{           \frac{n + 1}{\gamma (n-1)}     {\tilde{\mathcal D} }}
 \right\} 
\label{pdfstat}
\end{equation} 
 This measure is {\it not }  the canonical
 Boltzmann-Gibbs  distribution.
The crossover time  $t_c$ from  the  asymptotic distribution function
 for the undamped oscillator (\ref{pdfmult})
  to the stationary   measure in presence
 of damping (\ref{pdfstat}) is 
 $ t_c  =  \frac{1}{2 \gamma} \frac{n+1}{n-1}\, .$ 
 Thus,   
 Eq.~(\ref{pdfmult}), although derived
   for a non-dissipative system, is physically relevant:  it provides
  an intermediate time  asymptotic behavior  \cite{barenblatt}
 for the   P.D.F. before the stationary state is reached; 
  the  energy of the  system 
  is distributed  
 according to the time-dependent P.D.F. (\ref{pdfmult})
 for  times $t$ such that  $1 \ll t \ll t_c$ and then follows 
 the stationary law (\ref{pdfstat})  when $t \gg t_c$.

\begin{figure}[h]
\centerline{\includegraphics*[width=0.8\textwidth]{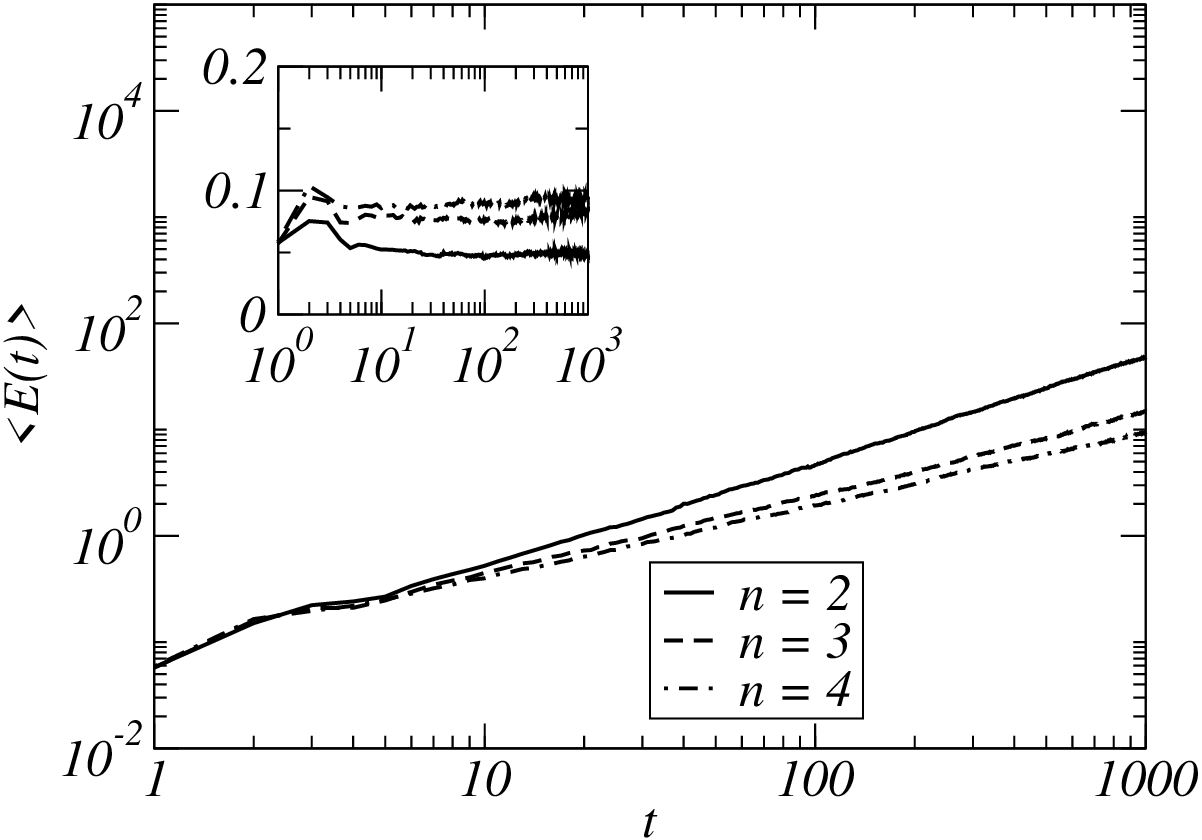}}
\caption{\label{fig:scaling:colored} 
Scaling behavior of $\langle E(t) \rangle$ for colored
Ornstein-Uhlenbeck multiplicative noise. 
Inset: the ratio $\langle E \rangle/t^{n/(2 (n-1))}$
is plotted vs. time.
Eqs.~(\ref{dyncol})-(\ref{OU}) are  integrated numerically for  
${\mathcal D} = 1$, $\tau = 1$,
with a timestep $\delta t$, and averaged over $10^3$ realizations for
$n=2$, $\delta t = 10^{-5}$;
$n=3$, $\delta t = 5 \, 10^{-6}$;
$n=4$, $\delta t = 2 \,  10^{-6}$.
}
\end{figure}

  \section{Colored  multiplicative  noise}
  
  We now consider   a system  without damping and submitted to a  
 multiplicative noise with  non-zero  correlation time $\tau$.
 In \cite{philkir1}, we  infered from  dimensional analysis
 that the white noise  scalings are  modified. 
 Here, we shall prove this conjecture by 
  calculating  the anomalous exponents from 
   a self-consistent approach. 

We study the equation:
 \begin{equation}
 \frac{\textrm{d}^2 }{\textrm{d} t^2}x(t) 
  + x(t)^{2n-1}  = \eta(t) \, x(t)   \,,
 \label{dyncol}
\end{equation}
where $\eta(t)$ is a colored, Ornstein-Uhlenbeck, 
Gaussian noise, obeying
  \begin{equation} 
 \frac{{\textrm d} \eta(t)}{{\textrm d} t} = -\frac{1}{\tau} \eta(t) + 
\frac{1}{\tau} \xi(t).
  \label{OU}
\end{equation} 
When $t, t' \gg \tau$, we find:
\begin{equation}
       \langle \eta(t)  \rangle =   0  \;\; \mathrm{and} \;\;
     \langle \eta(t) \eta(t') \rangle  =   
\frac{\mathcal D}{2 \, \tau}   \, {\rm e}^{-|t - t'|/\tau} \,.
   \label{deftau}
 \end{equation} 
 As stated earlier,
 Eqs.~(\ref{evoln1}) and (\ref{evoln})  remain  valid
 if $\xi(t)$ is replaced by $\eta(t)$.
  We assume that, in the long time limit, 
 $\Omega$ grows  algebraically with time with a scaling exponent $\alpha$,
 {\it i.e.}, 
 $   \Omega \sim t^{\alpha}   \, . $
 We deduce   from  Eq.~(\ref{evoln})  that $\phi \sim  t^{\nu}$
with $\nu =   \alpha + 1$.
Substituting  this scaling of $\phi$ in 
Eq.~(\ref{evoln1}), we obtain (leaving aside all proportionality
 constants): 
 \begin{equation}
    \Omega \sim  \int_0^t \rm{d}z  \, {\mathcal S}_n (z^{\nu})
 \, {\mathcal S}_n'(z^{\nu})  \, \eta(z)   \,.
 \label{eprcol}
\end{equation}
The asymptotic behavior of this expression can be estimated as in 
 \cite{philkir2} by discretizing time. This simplification
   does  not alter the critical
 exponents   but modifies only  the prefactors.
We  thus  replace the colored  Gaussian  noise $\eta$
  by a discrete random
 variable $\epsilon_k$ which 
  takes the values  $\pm \sqrt{  {\mathcal D}/(2 \tau)}$ randomly 
 during the time  interval $[k\tau \,  ,(k+1)\tau ] \, .$
 From Eq.~(\ref{eprcol}),  we deduce that
 \begin{equation}
 \langle \Omega^2 \rangle \sim  \sum_{k =0}^{t/\tau}
  \Bigg(  \int_{k\tau}^{(k+1)\tau} 
   \rm{d}z \; {\mathcal S}_n (z^{\nu})\,
   {\mathcal S}_n'(z^{\nu})  \Bigg)^2 \,  
   \sim \sum_{k =1}^{t/\tau}
  \frac{1}{\Big(k \tau \Big)^{2\nu-2}  } \sim  t^{3 - 2\nu} \, ,
   \label{Om2}
\end{equation}
   where the integral  is evaluated  by
integrating by parts and retaining only   the leading
  terms   \cite{philkir2}.
   Assuming that the variable $\Omega$ is not  multifractal,
 we also  have $\Omega ^ 2 \sim t^{2\alpha}. $ We then 
   deduce that  $  3 - 2\nu = 2 \alpha$. Since  
  $\nu = \alpha +1$,  we obtain 
  $  \alpha = {1}/{4} $ and   $\nu  = {5}/{4}$. Finally,
   the scaling exponents for colored noise  are 
\begin{equation}
          E      \sim  t^{\frac{n}{2(n-1)}}, \;\;\; 
          x      \sim  t^{\frac{1}{4(n-1)}} , \;\;\;
         \dot x  \sim t^{\frac{n}{4(n-1)} }  .
\label{scalingcolor}
\end{equation}
These predictions  are confirmed by  numerical  simulations
(see in Fig.~\ref{fig:scaling:colored}   the scaling behavior of
 the averaged energy).

\begin{figure}[ht]
\centerline{\includegraphics*[width=0.80\textwidth]{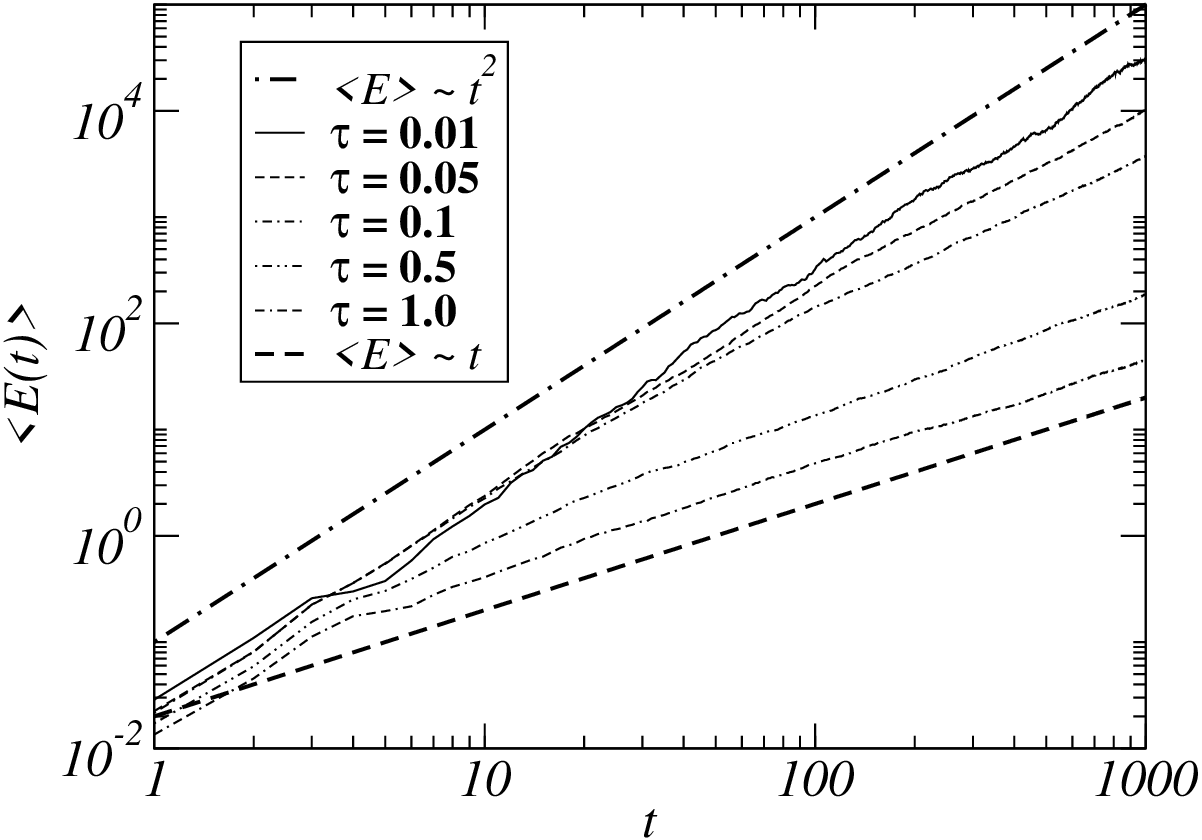}}
    \caption{\label{fig:crossover} 
Scaling behavior of $\langle E(t) \rangle$ for an oscillator in
a quartic potential ($n=3$) submitted to
Ornstein-Uhlenbeck multiplicative noise with various  values
 of the correlation time.
We find a crossover between the white noise
behavior observed at short times, $\, \langle E(t)\rangle  \propto  t^2\, ,$
and the colored noise behavior recovered at large enough times,  
$ \, \langle E(t) \rangle \propto t \, .$
}
\end{figure}

We observe that the  exponents are halved when the noise is colored.
Further, we expect that the colored noise scaling will be
observed for large enough time as soon as the correlation time
$\tau$ is nonzero.
This result may seem paradoxical at first sight  because
 it is often  wrongly  thought   that ``in the long  time limit,
 colored  noise  appears  white''. In fact
 the  period $T$ of the underlying  deterministic oscillator
   must be compared 
 with   the  correlation time  $\tau$ of the noise. 
 If   $\tau \ll T,$ the noise is uncorrelated over a period 
of  the deterministic oscillator  and  acts as if it were white:
 the scalings found in Eq.~(\ref{scalmult}) are satisfied.
 But,  for   $  T \ll \tau$, the noise remains  coherent
 over a period and its effect  is  perceptible
 only when it is considered over a large number of periods
 and hence  the diffusion  slows down: the scalings of
 Eq.~(\ref{scalingcolor}) now apply.  
  The  crossover between the two regimes
 is observed when $ T \sim  \tau$. The   deterministic
  period $T$  decreases   with the
 energy as $T \sim  E^{-\frac{n-1}{2n}}$ \cite{philkir1}.
 From Eq.(\ref{scalmult}), we have 
 $E^{-\frac{n-1}{2n}} \sim ({\mathcal D} t)^{-1/2}$. Therefore,
 the  crossover time  scales as $t_c \sim ({\mathcal D}\tau^2)^{-1}$.
As shown in Figure 3, numerical simulations confirm this scenario.
 Such a  crucial  relevance of the correlation time of the noise 
  was also found  
 in the  study  of  the bifurcation threshold  of the Duffing 
  oscillator  with multiplicative noise \cite{graham}.

 \section{Conclusion}

   A particle trapped in a confining potential and subject to
    multiplicative  noise undergoes anomalous diffusion before
 the dissipative time scale is reached. The critical
 exponents associated  with  this behaviour are  calculated
 exactly in the case of Gaussian white noise  and 
 an analytical expression for the  intermediate time asymptotics
 of the P.D.F. in phase space is derived. 

 Our approach relies  on the integrability of
  the underlying deterministic system: we 
 derive  exact stochastic equations in energy-angle variables and,  
  after  averaging out the phase variations,
 a  projected  dynamics for the energy  variable is obtained.
 Our analytical predictions   agree  perfectly  with  numerical  simulations.
 For colored  multiplicative  noise, the anomalous diffusion  exponents
  have been calculated
 in a self-consistent manner.
 We find that these  exponents are halved in presence of time correlations,
 however small the correlation time $\tau $ may be.

 We observe a crossover from the  white noise regime  at {\it short}
 times to the  colored noise scalings at {\it long} times.
 The  associated crossover time scales like $\tau^{-2}$ and corresponds
 to the regime when the period of the associated deterministic oscillator
 is of the order of $\tau $.
 In the colored noise  case, however, the
  averaging technique does not lead to conclusive results   because 
 the  noise itself  is  averaged out  to  the leading order. 
 A precise calculation in the case of colored Gaussian  noise 
 still remains to be done.

\end{document}